\documentclass[prb,showpacs,twocolumn]{revtex4}

\usepackage{graphicx}
\usepackage[latin1]{inputenc}
\usepackage{amsmath}

\newcommand{\bm}[1]{\boldsymbol{#1}}
\newcommand{\eq}[1]{Eq.\,(\ref{#1})}

\newcommand{\SKIP}[1]{}

\newcommand{\fig}[1]{Fig.\,\ref{#1}}
\newcommand{\sect}[1]{Sec.\,\ref{#1}}
\newcommand{\R}{{\bm R}}
\newcommand{\intra}{{(\mbox{\small c})} }
\newcommand{\inter}{{(\mbox{\small i})}}
\newcommand{\tr}{{\rm tr}\,}

\newcommand{\onlref}[1]{Ref.\,\onlinecite{#1}}
\newcommand{\nn}[1]{\langle #1\rangle}
\newcommand{\sij}[1]{\sum_{\langle ij#1\rangle}}

\begin{document}

\title{Single-particle spectral function of quarter-filled ladder systems}

\author{M. Aichhorn}
\author{E. Ya. Sherman}
\author{H. G. Evertz}
\affiliation{Institut f\"ur Theoretische Physik - Computational Physics,
Technische Universit\"at Graz,
Petersgasse 16, A-8010 Graz, Austria}


\begin{abstract}

We study the single-particle properties of quarter-filled ladder systems such
as $\alpha^\prime$-NaV$_2$O$_5$ by
means of a recently developed generalization of the variational cluster
perturbation theory to extended Hubbard models. 
We find a homogeneous antiferromagnetic
insulating phase for nearest-neighbor repulsions smaller than a critical
value, without any metallic phase for small repulsions.
Different from C-DMFT and LDA considerations,
the inclusion of diagonal hopping within a ladder has little effect on
the bonding bands, while flattening and shifting the antibonding bands.
In the low-temperature charge-ordered phase,
the spectrum depends on whether the ordering is driven by the Coulomb repulsion
or by the coupling to a static lattice distortion.
The small change of the experimentally observed gap upon charge ordering
implies that the lattice coupling plays an important role in this ordering.
Inter-ladder coupling is straightforward to include within our method.
We show that it has only a minor effect on the spectral function. 
The numerically calculated spectra show good
agreement with experimental angle-resolved photo-emission data.

\end{abstract}


\maketitle

\section{Introduction}

In recent years low-dimensional strongly-correlated systems have been the subject
of many experimental and theoretical studies due to their fascinating 
properties such as the occurrence of ordered patterns of the ion
charges. A compound in this class of materials is the low-dimensional
vanadium bronze $\alpha^\prime$-NaV$_2$O$_5$. Although known for many
years\cite{CaGa75} it has attracted considerable attention in recent years
because of a very interesting low-temperature phase. The compound exhibits a
spin-Peierls-like transition at $T_c\approx 35$\,K accompanied by the opening
of a spin gap.\cite{IsUe96} At the same\cite{Popova02} or slightly higher
temperature charge ordering takes place. Different from the first X-ray
investigations, recent studies\cite{Schn98,Smolinski98,Meet98} at room
temperature showed a disordered state 
with equivalent valence 4.5 for all vanadium ions. Below the phase transition
point, NMR-studies\cite{Ohama99} gave two different valences for the ions, a
clear evidence for the formation of a charge-order pattern below $T_c$. Since
one $d_{xy}$ electron is shared by two V sites in a V-O-V rung, the ordering
occurs as a static charge disproportion $\delta$ between the V ions, yielding
charges $4.5\pm\delta$ with a zig-zag pattern of $\delta$'s. Since 
the crystal environment of the vanadium ions is asymmetric, the $d_{xy}$
electrons are coupled to the lattice via a strong Holstein-like
electron-phonon interaction.\cite{Sherman99} This results in a static lattice
distortion below the charge ordering transition temperature, where the ion
displacements from their positions in the high-temperature phase is of the
order of $0.05$\,\AA\ as observed in X-ray diffraction
experiments.\cite{Luedecke99}

Although the crystal structure of $\alpha^\prime$-NaV$_2$O$_5$ is composed of
nearly decoupled two-dimensional layers that consist of coupled two-leg
ladders, spin-susceptibility measurements\cite{IsUe96} revealed that the
system can be reasonably well 
described by a one-dimensional Heisenberg model. This behavior could be
explained by realizing that the molecular-orbital state on a rung occupied by
one electron is a key element of the electronic structure,\cite{Horsch98}
yielding quasi 1D magnetic exchange couplings. In addition angle-resolved
photo-emission spectroscopy (ARPES), performed in the disordered
high-temperature phase showed  
quasi 1D band dispersions of the vanadium $3d$
bands,\cite{Kobayashi98,Damascelli04} and it was argued that 
spin-charge separation should be present in this system.\cite{Kobayashi99}

Previous
studies\cite{Riera99,Clay03,Sherman01,Aichhorn04} revealed that the
electron-phonon coupling is very important for the phase transition in
quarter-filled ladder compounds. For this reason we study a model Hamiltonian
that includes the coupling of the $d_{xy}$ electrons to the lattice. The
relevant parameters for the study of lattice effects can be obtained
from experiments\cite{Fischer99} (phonon frequencies), and from
first-principle calculations (lattice force constants and 
electron-phonon coupling).\cite{Smolinski98,Spitaler03}

Static and dynamic properties of quarter-filled ladder compounds without
coupling to the lattice have been studied intensively in the past using
different methods like 
mean-field approaches,\cite{Seo98,Thalmeier98,Mostovoy00} exact
diagonalization (ED) of small clusters,\cite{Cuoco99,Hubsch01,Aichhorn02,Riera99_2}
density-matrix renormalization-group (DMRG),\cite{Vojta01} cluster dynamical
mean-field theory (C-DMFT),\cite{Mazurenko02} and bosonization
and renormalization-group techniques.\cite{Orignac03} Recently the influence
of the lattice coupling on the charge-ordering transition was investigated by
employing ED methods.\cite{Aichhorn04}

In this paper we study the single-particle spectral function of the compound
$\alpha^\prime$-NaV$_2$O$_5$, which can be directly related to the ARPES
experiments, by applying the recently proposed variational 
cluster perturbation theory (V-CPT).\cite{Potthoff03_2} This method
is a combination of the cluster perturbation theory\cite{Gros93} and the
self-energy functional approach (SFA),\cite{Potthoff03} which provides
results for the infinite lattice and allows to
study symmetry-broken phases. It was used with success for the
investigation of the magnetic ground-state properties of the two-dimensional
Hubbard model.\cite{Dahnken03} For Hamiltonians including offsite Coulomb
interactions an extension of this theory has been developed,\cite{Aichhorn04_2}
which turned out to give very accurate results for the one and
two-dimensional extended Hubbard model.

The paper is organized as follows. In \sect{sec:model} we introduce the model
Hamiltonian and give a short description of the V-CPT
method. \sect{sec:results_1l} and \sect{sec:results_coupled} includes our
results for single and coupled ladders, respectively, and we finally draw our
conclusions in \sect{sec:conclusion}.

\section{Model and method}\label{sec:model}

\begin{figure}[t]
  \centering
  \includegraphics[width=0.45\textwidth]{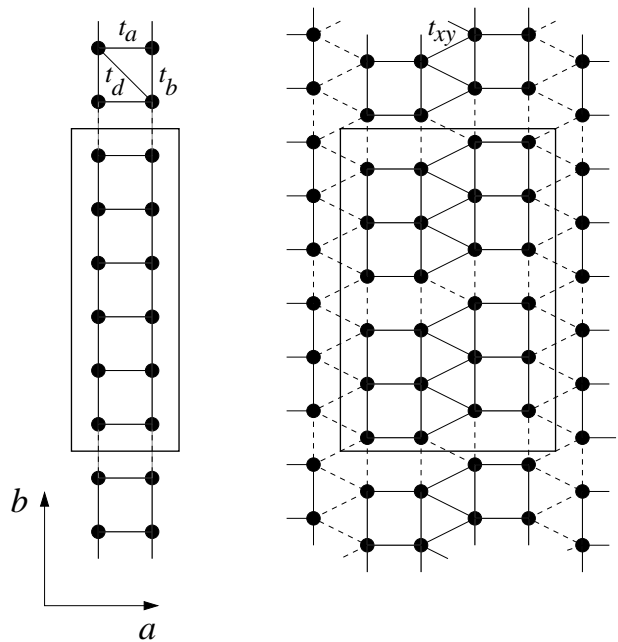}
  \caption{\label{fig:clusters}%
    Clusters used for the V-CPT calculations. The diagonal hopping $t_d$ is
    indicated only once, but is present equivalently between other
    sites. Decoupled bonds treated perturbatively are marked by
    dashed lines, and the boxes show the clusters of finite size. Left:
    Single ladder with $6\times2$ cluster. Right: Super cluster consisting of
    two 12 site clusters.}
\end{figure}

On a microscopic level, $\alpha^\prime$-NaV$_2$O$_5$ can be described by an
extended Hubbard model (EHM). In order to take into account lattice effects
we further extend this well-known model by a Holstein-like electron-phonon
coupling, which a recent LDA study\cite{Spitaler03} showed to be especially
important, yielding the model
\begin{equation}\label{eq:hamiltonian}
  H = H_{\rm EHM}+H_{\rm l}+H_{\rm e-l},
\end{equation}
with $H_{\rm EHM}$ the EHM Hamiltonian, $H_{\rm l}$ the contribution of the
lattice, and $H_{\rm e-l}$ the Holstein coupling. These
terms are given by
\begin{subequations}
\begin{align}
    H_{\rm EHM}=&-\sum_{\langle ij\rangle,\sigma}t_{ij}\left(c_{i\sigma}^\dagger
    c_{j\sigma}^{\phantom{\dagger}}+\mbox{H.c.}\right)\label{hehm}\nonumber\\
    &+U\sum_in_{i\uparrow}n_{i\downarrow}+\sij{}V_{ij}n_in_j,\\
    H_{\rm l}= &\kappa\sum_i\frac{z_i^2}{2},\label{eq_2b}\\
    H_{\rm e-l}=&-C\sum_iz_in_i,\label{eq_2c}
\end{align}
\end{subequations}  
where $\nn{ij}$ connects nearest-neighbor bonds, and $t_{ij}$ is the
corresponding hopping matrix element. In \fig{fig:clusters} the lattice
structure and the hopping processes used in this study are shown. The most
commonly used set for these matrix elements 
is $t_a=0.38$\,eV, $t_b=0.18$\,eV, and $t_{xy}=0.012$\,eV, and was obtained by fitting 
the LDA bands.\cite{Smolinski98} A recent study\cite{Spitaler03} gave similar
parameter values. By including the additional hopping term $t_d$ in a massive
downfolding procedure, Mazurenko {\it et.\,al}\cite{Mazurenko02} found similar
values for $t_a$ and $t_{xy}$, but the values $t_b=0.084$ and $t_d=0.083$
differ considerably from previous studies. In the present study we set $t_a$ as the
energy unit and fix $t_b=0.5$, except for \sect{subsec:diagonalhop}, where we
study the spectral function including the
hopping term $t_d$. The onsite Coulomb interaction is set to $U=8$
throughout the paper, in accordance with band-structure
calculations,\cite{Spitaler03} and the inter-site Coulomb interaction
$V_{ij}$ is 
treated as a free parameter of the system, since the determination of a
proper value within first-principle calculations is very difficult.

The lattice distortions are given in units of 0.05\,\AA, since the ion
displacements in the ordered phase are of this order of magnitude.
The electron-phonon coupling $C$ and the lattice rigidity $\kappa$ were 
determined by first-principle calculations\cite{Spitaler03} yielding
$C=0.35$ and $\kappa=0.125$ in these units. We restrict our
investigations to static distortions and neglect dynamical phonon effects,
similar to parts of \onlref{Aichhorn04}. Moreover, as discussed below, we use
a staggered zig-zag configuration for the $z_i$, as observed
experimentally.\cite{Luedecke99} 

The method we use in this paper for the calculation of the single-particle
spectrum is the variational cluster perturbation theory for an
EHM.\cite{Aichhorn04_2} The main idea is to 
decouple the lattice into clusters of finite size as depicted in
\fig{fig:clusters}, yielding the Hamiltonian 
\begin{equation}\label{eq:ham_decoupled}
  H=\sum_\R\left[H_0^\intra(\R)+H_1(\R)\right]+\sum_{\R,\R^\prime}H_0^\inter(\R,\R^\prime),
\end{equation}
where $\R$ denotes the individual clusters. The first sum consists of
decoupled intra-cluster Hamiltonians with interaction part $H_1(\R)$, and the
second sum gives the coupling between clusters, which must be of
single-particle type and is of the general form
\begin{equation}\label{eq:interclhop}
  H_0^\inter(\R,\R^\prime)=\sum_{a,b}T_{a,b}^{\R,\R^\prime}c_{\R,a}^\dagger
  c_{\R^\prime,b}^{\phantom{\dagger}}.
\end{equation}
The indices $a$ and $b$ are orbital indices within a cluster.

In the case of the EHM this decoupling into clusters yielding
\eq{eq:ham_decoupled} cannot be done in a straightforward way, since the
Coulomb interaction on the decoupled bonds is of two-particle type. In order
to get a Hamiltonian of the form \eq{eq:ham_decoupled} it is necessary to do a
mean-field decoupling of the Coulomb-interaction terms on inter-cluster bonds,
which is described in detail in \onlref{Aichhorn04_2}. The coupling of the
clusters is then again of single-particle type, \eq{eq:interclhop}, but in
addition this mean-field approximation introduces onsite potentials
$\lambda_i$ on the cluster boundary, which correspond to the average
electronic densities on site $i$ of the cluster boundary. Let us stress at
this point that the $\lambda_i$ are not variational parameters 
in the sense of V-CPT but external parameters to the Hamiltonian
\eq{eq:ham_decoupled} in its mean-field decoupled form, entering term
$H_0^\intra$.    

After the mean-field decoupling of inter-cluster bonds, we can apply the V-CPT
to the Hamiltonian \eq{eq:ham_decoupled}.
Obviously the Hamiltonian is invariant under the transformation
\begin{equation}\label{eq:transf}
  \begin{split}
  H_0^\intra(\R)&\to H_0^\intra(\R)+\mathcal{O}(\R)\\
  H_0^\inter(\R,\R^\prime)&\to H_0^\inter(\R,\R^\prime)
  -\delta_{\R,\R^\prime}\mathcal{O}(\R),
  \end{split}
\end{equation}
with an arbitrary single-particle operator
\begin{equation}\label{eq:add_onepartop}
  \mathcal{O}(\R)=\sum_{a,b}\Delta_{a,b}\, c_{\R,a}^\dagger
  c_{\R,b}^{\phantom{\dagger}}.
\end{equation}
This transformation allows the study of symmetry-broken phases by the
inclusion of fictitious variational symmetry-breaking fields, which do {\em
  not} change the overall Hamiltonian, \eq{eq:ham_decoupled}, but instead just
rearrange it. They are therefore different in character from the external
lattice distortions, \eq{eq_2b}, and also from the external mean-field
parameters $\lambda_i$, which directly enter the Hamiltonian. For our model,
symmetry-breaking fields corresponding to charge order and to magnetic order
can be important. We will later see that we can omit a charge-order
symmetry-breaking field. A symmetry-breaking field for magnetic order will be
considered in \sect{subsec:disordered}.

After introducing the external mean-field parameters $\lambda_i$ and the
variational 
parameters $\bm \Delta =\Delta_{a,b}$, they have to be determined in a proper
way, which is done within the framework of the SFA.
It provides a unique way to calculate the grand potential
of a system by using dynamical information of an exactly solvable reference
system, which is in our case the decoupled cluster. This grand potential is
parameterized as a function of 
the external mean-field parameters $\lambda_i$ and the
variational parameters $\bm \Delta =\Delta_{a,b}$.  
The functional form of the grand potential is taken from 
Refs.~\onlinecite{Potthoff03,Aichhorn04_2}: 
\begin{align}\label{eq:sfa_omega}
  \Omega(\bm \Delta,\lambda_i)&=\Omega^\prime(\bm \Delta,\lambda_i)\nonumber\\
  &+T\sum_{\omega_n,\bm{q}}\tr \ln\frac{-\bm
    1}{\bm{G}_{\bm{q}}^{(0)}(\lambda_i,i\omega_n)^{-1}-\bm\Sigma(\bm
    \Delta,\lambda_i,i\omega_n)} \nonumber\\
  &-LT\sum_{\omega_n}\tr\ln(-\bm G^\prime(\bm \Delta,\lambda_i,i\omega_n)),
\end{align}
where $\Omega^\prime(\bm \Delta,\lambda_i)$ is the grand potential of the
decoupled 
clusters, $\bm{G}_{\bm{q}}^{(0)}(\lambda_i,i\omega_n)$ is the non-interacting
Green's 
function of the original infinite-lattice problem after mean-field decoupling
of 
inter-cluster Coulomb interactions, $\bm\Sigma(\bm
\Delta,\lambda_i,i\omega_n)$ is the cluster self energy, $\bm
G^\prime(\bm \Delta,\lambda_i,i\omega_n)$ the cluster Green's
function and $L$ denotes the number of clusters. All cluster properties can
easily be calculated by the Lanczos algorithm. The sum over Matsubara frequencies in
\eq{eq:sfa_omega} is evaluated by a continuation to the real frequency axis,
$i\omega_n \to \omega+i0^+$, yielding an integral from minus infinity to the
chemical potential $\mu$, determined below. Note that the Hamiltonian,
\eq{eq:hamiltonian}, does not involve $\mu$ in any way.

The general procedure to determine the $\lambda_i$ and the values of the
variational parameters $\Delta_{a,b}$ is the following. 
First one has to distinguish between the external parameters $\lambda_i$ and
the variational parameters $\Delta_{a,b}$. For the latter ones the general
variational principle of the SFA says that $\Omega$, \eq{eq:sfa_omega}, must
have a stationary point with respect to $\Delta_{a,b}$, but the SFA does not
provide any information on the second derivative. That means that this
stationary point can be a maximum, minimum, or a saddle point. 
The situation is different for the external parameters. Since the $\lambda_i$
are mean-field parameters, one has to look for a minimum of $\Omega$ with
respect to the $\lambda_i$. Finding the minimum in the grand potential is
equivalent to a self-consistent solution for the $\lambda_i$, as shown
in the appendix of \onlref{Aichhorn04_2}. 

In practice we used the following procedure.
For a given value of
the $\lambda_i$ one has to find the stationary point of $\Omega(\bm
\Delta,\lambda_i)$ with respect to $\bm \Delta$, yielding a function
$\Omega=\Omega(\lambda_i)$. The proper value of the external parameters
$\lambda_i$ is then given by the minimum of this function. 

In \sect{subsec:ordered} we study the effect of the lattice distortions
$z_i$, and the optimal distortions are determined by the minimum ground-state
energy. From a technical point of view these
distortions are treated on the same level as the mean-field parameters
$\lambda_i$, since they are external parameters to the Hamiltonian,
\eq{eq:hamiltonian}, as well. In 
this case on has a function $\Omega(\bm \Delta,\lambda_i,z_i)$. Again, for
each pair $\lambda_i,z_i$ one looks for the stationary point with respect to
$\bm \Delta$, and the proper choice of $\lambda_i$ and $z_i$ is then given by
the minimum of the function $\Omega(\lambda_i,z_i)$.

The single-particle Green's
function is then calculated by
\begin{equation}\label{eq:cpt_gq}
  \bm{G}_{\bm{q}}(\omega)=\left[\bm
  G^\prime(\omega)^{-1}-\bm{T}_{\bm{q}}\right]^{-1}
\end{equation}
with the Fourier-transformed matrix elements $T_{\bm{q},a,b}$.\cite{Gros93}
After 
applying a residual Fourier transformation\cite{Gros93} one finally obtains
the fully 
momentum-dependent Green's function $G(\bm k,\omega)$ for the infinite size
system. Note that the external
parameters ($\lambda_i,z_i$) are only present in the calculation of
$\bm{G}^\prime(\omega)$, whereas the variational parameters $\bm \Delta$ also
enter $\bm{T_q}$, see
Eqs.\,(\ref{eq:interclhop},\ref{eq:transf},\ref{eq:add_onepartop}).  

Since calculations are not done at half filling, the chemical potential $\mu$
is not known a priori. However, the knowledge of $\mu$ is important for the 
evaluation of the grand potential, as discussed above.
One can calculate $\mu$ from the condition
\begin{equation}\label{eq:mucpt}
  n=\frac{2}{L}\sum_{\bm k}\int_{-\infty}^\mu\!\!\!{\rm d}\omega
  A({\bm k},\omega), 
\end{equation}
where the spectral function $A(\bm k,\omega)$ is given by
\begin{equation}
  A({\bm k},\omega) = -\frac{1}{\pi}{\rm Im}\, G(\bm k,\omega+i\eta)
\end{equation}  
and $\eta$ is a small Lorentzian broadening. This amounts to a self-consistent
procedure, since for the calculation of $\mu$ the Green's function
$G(\bm k, \omega)$ is needed, and for the determination of $G(\bm k, \omega)$
one has 
to know $\mu$. This cycle can be avoided as follows. One can infer the
chemical potential directly from the energies of the excited states obtained
by the ED. An approximate value for the chemical potential is
given by 
\begin{equation}\label{eq:mued}
  \mu_{\rm ED}=\frac{E_{\rm min}^{\rm IPES} + E_{\rm max}^{\rm PES}}{2},
\end{equation}
with $E_{\rm min}^{\rm IPES}$ the minimal energy of inverse-photo-emission (IPES)
states and $E_{\rm max}^{\rm PES}$ the maximum energy of photo-emission (PES)
states. This value only weakly depends on the mean-field and
variational parameters. As discussed below, in all our calculations we found a
well established gap between 
the PES and IPES states yielding a constant density $n$ in a reasonably large
neighborhood of the physical chemical potential $\mu$, in agreement with
quantum Monte Carlo calculations.\cite{Gabriel04} Therefore $\mu_{\rm
  ED}$ gives a reasonable approximation for our calculations.

For the details of this method and the
calculation of the grand potential we refer the reader to
Refs.~\onlinecite{Gros93,Potthoff03_2,Dahnken03,Aichhorn04_2} and references therein. 

\section{Results for single ladders}\label{sec:results_1l}

\subsection{Critical coupling}

We start our investigations with decoupled ladders, i.e., $t_{xy}=0$ and
$V_{xy}=0$ (\fig{fig:clusters}). Before we turn to the spectral function, we study the
charge-ordering transition as a function of $V=V_a=V_b$. Let us first look on
the effect of the mean-field decoupling of the nonlocal Coulomb interactions
across cluster boundaries as described in \sect{sec:model}.
For this purpose we consider
the EHM without coupling to the lattice in the limit of exactly one electron
per rung and $t_b=0$, i.e. without hopping between rungs. In mean-field
approximation this case results in a 
second-order phase transition between a disordered state and a 
zig-zag ordered state at a critical interaction of $V_c^{\rm MF}=1.0$. On the other
hand this case is exactly solvable by a mapping to an Ising
model in a transverse field,\cite{Mostovoy00} yielding a critical interaction
of $V_c^{\rm exact}=2.0$.\cite{Lieb61} Thus we expect strong
mean-field effects, since in this special limit we found $V_c^{\rm exact}=2
V_c^{\rm MF}$. Since it can be assumed that a finite value of the hopping
between adjacent rungs $t_b$ weakens the charge ordering, the actual critical
value $V_c$ is presumably located slightly above 2.0 when $t_b$ is included.

For single ladders at quarter filling one has only two different values for the
mean-field parameters $\lambda_i$, namely $\lambda_A$ on sublattice $A$ and
$\lambda_B$ on 
sublattice $B$, which correspond to the left and right side of the rung on the
cluster boundary, respectively, see \fig{fig:clusters}. In order to reduce the
number of mean-field parameters, we set $\lambda_A=\langle n\rangle
+\delta$ and $\lambda_B=\langle n\rangle-\delta$, with the average density
fixed to $\langle n\rangle =0.5$. This yields only one mean-field parameter $\delta$
instead of two parameters, $\lambda_A$ and $\lambda_B$.  

For second-order phase transitions it can in addition
be important to rearrange the Hamiltonian by means of a fictitious staggered
chemical potential as a 
variational parameter.\cite{Aichhorn04_2} This field is included by adding and
substracting (\eq{eq:transf}) the single-particle operator $\mathcal O(\R)$,
\eq{eq:add_onepartop}, with 
\begin{equation}\label{eq:staggeredfield}
  \Delta_{a,b}=\varepsilon\delta_{a,b}e^{i\bm{Q}\bm{r}_a},
\end{equation}
where $\varepsilon$ denotes the variational parameter, $\bm{r}_a$ is the
lattice vector of site $a$, $\delta_{a,b}$ is the Kronecker-$\delta$ for sites
$a,b$, and $\bm{Q}=(\pi,\pi)$. Initial calculations
showed that in the present case the inclusion of such a field does not have
any significant effect and the relative change in $\Omega$ is at most of the
order of $10^{-4}$. Although the staggered field \eq{eq:staggeredfield} is the
Weiss field associated with the CDW, its effect is very small, since the
symmetry is already broken by the mean-field decoupling at the boundaries. 
We also considered a staggered chemical potential similar to
\eq{eq:staggeredfield}, but which is nonzero only on the
cluster boundaries. In the spirit of V-CPT this corresponds to include the
mean-field parameter $\delta$ also in the set of variational parameters $\bm
\Delta$. Not surprisingly, the effect of this field was even smaller than for the field
\eq{eq:staggeredfield}. For this reason all further calculations have been done
without a staggered chemical potential.

\begin{figure}[t]
  \centering
  \includegraphics[width=0.45\textwidth]{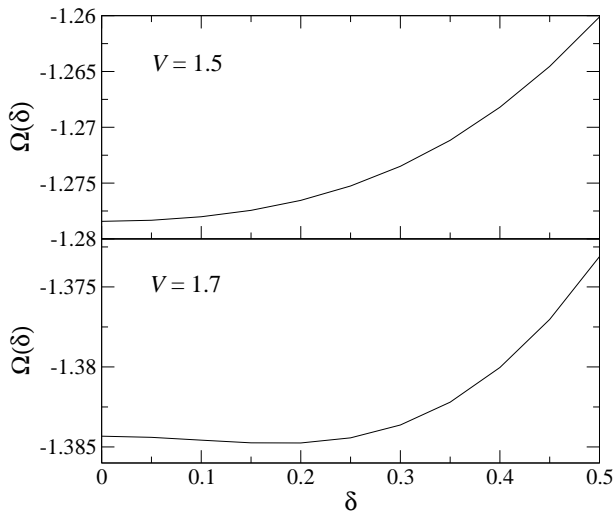}
  \caption{\label{fig:omegavsdelta}%
    Grand potential $\Omega(\delta)$ as a function of the mean-field
    parameter $\delta$ with a $6\times 2$ cluster serving as reference
    system, and without coupling to the lattice. Upper panel: $V=1.5$. Lower
    panel: $V=1.7$.}  
\end{figure}

We now discuss our results, setting $t_b=0.5$. We first discuss the case
without coupling to the lattice. Finite lattice coupling will be considered in
\sect{subsec:ordered}. In order to determine the order of the transition
within the framework of V-CPT it is sufficient to calculate the grand
potential $\Omega$, \eq{eq:sfa_omega}, as a function of the mean-field 
parameters.\cite{Aichhorn04_2} 
\fig{fig:omegavsdelta} shows the dependence of $\Omega(\delta)$ on the
mean-field parameter $\delta$ calculated with a $6\times 2$ cluster as
reference system. One can see that the
system undergoes a continuous phase transition,\cite{Aichhorn04_2} which is
located between $V=1.5$ and $V=1.7$, since the minimum of $\Omega(\delta)$
shifts from $\delta=0$ to a finite value. This value for the 
critical interaction is considerably smaller than the above mentioned value
of the analytical solution, but the agreement is much better
than the result of the purely mean-field calculation, $V_c^{\rm MF}=1.0$.

\begin{figure}[t]
  \centering
  \includegraphics[width=0.47\textwidth]{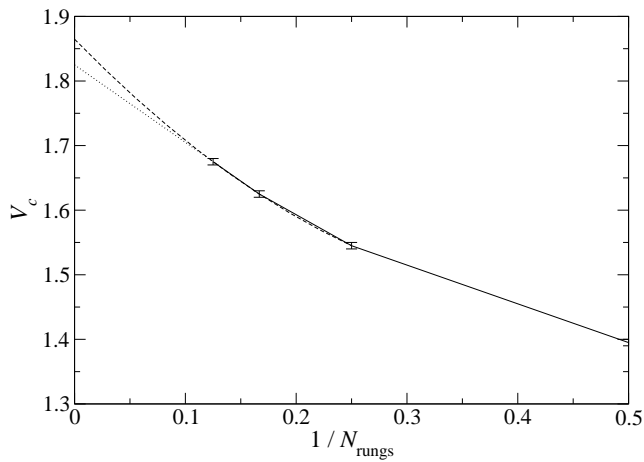}
  \caption{\label{fig:vcrit}%
    Finite-size dependence of the critical Coulomb interaction
    $V_c$ without lattice coupling. Error bars are due to the finite step
    $\Delta V=0.01$ in the calculations. Dotted line: Linear extrapolation of
    the 8 and 6 rung cluster. Dashed line: Quadratic extrapolation of the 8,
    6, and 4 rung cluster.} 
\end{figure}

In order to study the finite-size dependence of the critical Coulomb
interaction we performed calculations on clusters of different length, and the
results are depicted in \fig{fig:vcrit}. The steps in $V$ in our calculations
were $\Delta V=0.01$, which results in error bars of $\Delta V_c=0.005$. As
expected, $V_c$ is 
strongly finite-size dependent. From \fig{fig:vcrit} we can 
expect that for larger cluster sizes the critical interaction $V_c$ increases
further and reaches the expected value of slightly above 2.0, but for a more
sophisticated finite-size scaling our cluster sizes are too
small. Nevertheless it is possible to study the spectral function both in the
disordered and the ordered phase.
Since the
calculations for the $8\times 2$ ladder are very time consuming, all
single-ladder spectra presented in this paper have been determined with a
$6\times 2$ ladder as reference system.

\subsection{Disordered phase}\label{subsec:disordered}

\begin{figure}[t]
  \centering
  \includegraphics[width=0.45\textwidth]{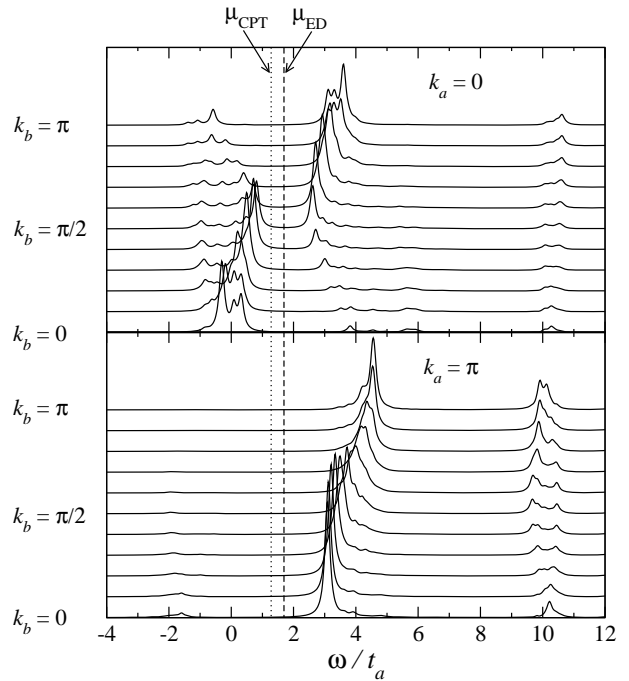}
  \caption{\label{fig:akw_1l_v13}%
    Single-particle spectral function $A({\bm k},\omega)$ calculated with a
    $6\times 2$ cluster in the disordered phase at $V_a=V_b=1.3$. Top panel:
    Momentum $k_a=0$ perpendicular to the ladder. Bottom panel:
    $k_a=\pi$. The dashed line marks the chemical potential calculated by
    \eq{eq:mued}, the dotted line marks the result obtained from \eq{eq:mucpt}.}
\end{figure}

We start our investigations of the spectral function with the disordered
high-temperature phase. Since $\alpha^\prime$-NaV$_2$O$_5$ may be near a quantum
critical point between ordered and disordered phase, we choose the nearest
neighbor interaction to be slightly below the critical value. 
We set $V=V_a=V_b=1.3$, $t_b=0.5$, and we do not include diagonal hopping,
i.e. $t_d=0$. The result of this calculation is shown in
\fig{fig:akw_1l_v13}. An additional Lorentzian broadening of $\eta=0.1$ has been
used for all spectra shown in this paper. The dashed vertical line marks $\mu_{\rm ED}$
calculated from \eq{eq:mued}, and the dotted line denotes $\mu_{\rm CPT}$
determined from the condition \eq{eq:mucpt}. For the latter quantity the sum
over momentum vectors had to consist of about 80 vectors in order
to get a well converged result. It is easy to see that $\mu_{\rm ED}=1.71$ lies
exactly in the middle of the gap, whereas $\mu_{\rm CPT}=1.23$ is located at its
lower boundary. But since there are no in-gap states both values
of $\mu$ give approximately the same average density $n$, and the
ground-state energy $E_0=\Omega +\mu N$ hardly depends on whether we use
$\mu_{\rm ED}$ 
or $\mu_{\rm CPT}$. These facts confirm that our approximation to use
$\mu=\mu_{\rm ED}$ as chemical potential gives correct results, and in
addition the numerical effort for this procedure is much less than for the
above described self-consistent determination of $\mu$.

\begin{figure}[t]
  \centering
  \includegraphics[width=0.45\textwidth]{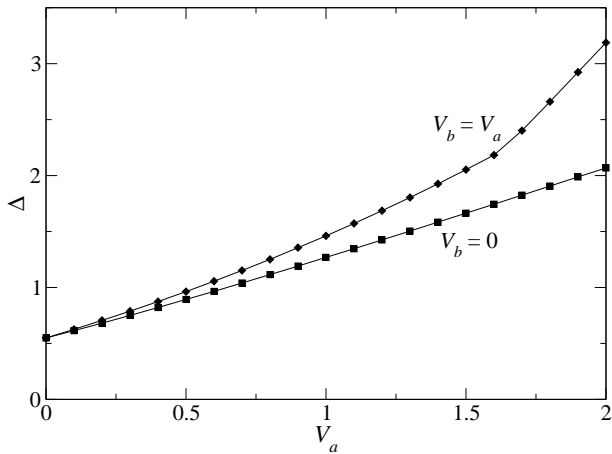}
  \caption{\label{fig:gap}%
    Gap $\Delta$ in the spectral function as a function of $V_a$. Squares:
    $V_b=0$. Diamonds: $V_b = V_a$.}
\end{figure}

As one can easily see in \fig{fig:akw_1l_v13}, the spectral function exhibits
a well defined gap around the chemical potential, a clear indication of
insulating behavior. In order to check if the insulator is only stable above
some critical inter-site Coulomb interaction, we calculated the gap $\Delta$ at
$(k_a,k_b)=(0,\pi/2)$ as a function of the intra-rung interaction $V_a$. We
studied two cases with $V_b=0$ and $V_b=V_a$, respectively, and the results
are shown in \fig{fig:gap}. Note that for $V_b=0$ no mean-field decoupling is
needed, since there are no interaction bonds between different clusters.
At $V_a=0$, where both cases are equivalent, we found a finite value of the
gap, $\Delta\approx 0.55$. We checked the finite size dependence by
calculating the gap on a $4\times 2$ cluster giving $\Delta\approx 0.59$. By
applying a linear $1/N_{\rm rungs}$ extrapolation to $N_{\rm rungs}=\infty$
one gets $\Delta\approx 0.47$, indicating that the curves in \fig{fig:gap}
somewhat overestimates the  
value of the gap for the infinite ladder. Nevertheless we conclude from our
calculations that the system is insulating already for small values of $V_a$.
This is consistent with DMRG calculations,\cite{Vojta01} where for $t_a>t_b$
a homogeneous insulating phase has been found 
for $V=0$. The behavior of the spectral function is also in agreement with ED
calculations on small clusters for $V=0$, where for large enough $t_a$ an
insulating state has been found.\cite{Riera99_2} Similar results have been

obtained by Kohno\cite{Kohno97} for the $U=\infty$ Hubbard ladder.

In the case $V_b=0$, which means that there is no Coulomb interaction between
adjacent rungs, we found that $\Delta$ increases linearly with $V_a$. For
$V_b=V_a$ the gap is slightly larger and the deviation increases with
increasing $V_a$. Here, with a $6\times 2$ cluster as reference system, the
system starts to order at
$V_c\approx 1.625$, which results in the kink in $\Delta$ around this
critical value. Note that for $V_b=0$ such a phase transition is not
possible.

Let us now discuss the spectral features for $k_a=0$ as shown in
\fig{fig:akw_1l_v13}. The spectral function looks very similar to
that of the half-filled one-dimensional Hubbard model with a totally filled
lower and an empty upper band. Different from the 1D Hubbard model the
gap between these two bands is not only determined by the onsite interaction $U$,
but mainly by the intrarung interaction $V_a$, as discussed above.
At ${\bm k}=(0,0)$ one can see signatures of
spin-charge separation, where the band is split into a low energy spinon
band (at approximately $\omega-\mu\approx -1.5$) and a holon band at slightly
higher energy ($\omega-\mu\approx -2.0$). This splitting has not been seen
directly in experiments,\cite{Kobayashi98,Damascelli04} since it is small and 
temperature effects did not allow a high enough experimental
resolution. However, by studying the temperature dependence of ARPES spectra,
it was argued that subtle spectral-weight redistributions can be related to
spin-charge separation.\cite{Kobayashi99}
 Some
spectral weight can also be found at very high energies of about $(\omega -
\mu)\approx 8.5$, which is close to the onsite energy $U=8$ and can thus be
related to doubly-occupied sites.

Infrared (IR) experiments
probe transitions near the 
$\Gamma$ point, that is between even $(0,0)$ and odd $(0,\pi)$ states in the
language of single ladders. From \fig{fig:akw_1l_v13} one can extract an
excitation energy of roughly $3t_a$, which is in good agreement with the
experimentally found 1\,eV absorption peak.\cite{Presura00}

\begin{figure}[t]
  \centering
  \includegraphics[width=0.45\textwidth]{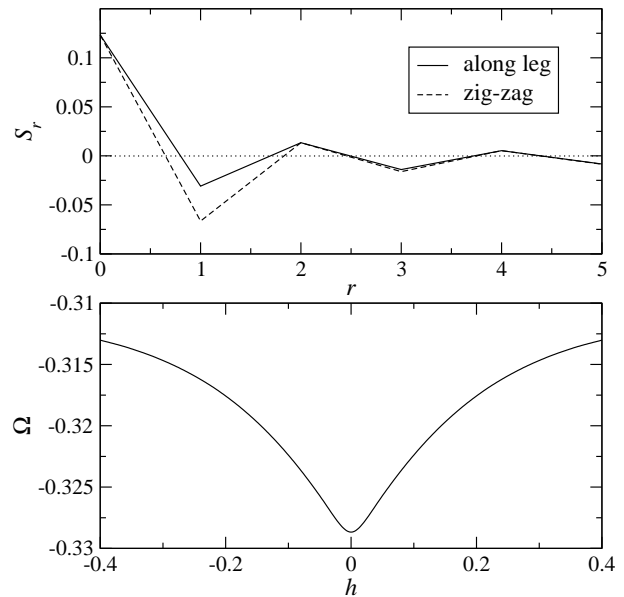}
  \caption{\label{fig:magn}%
    Magnetic properties of a single ladder at $V_a=V_b=1.3$. Upper panel:
    Spin correlation function $S_{r}=\langle S_1^z S_{1+r}^z\rangle$ calculated
    on an isolated $6\times 2$ cluster. Lower panel: Grand potential $\Omega$
    as a function of the strength of the fictitious field \eq{eq:magnfield}.}
\end{figure}

The lower and upper band disperse with period $\pi$ indicating a
doubling of the unit cell in real space, similar to the 1D Hubbard model. In
order to determine the origin of this doubling we have calculated the real-space spin
correlation function $S_{r}=\langle S_1^z S_{1+r}^z\rangle$ within the cluster by exact
diagonalization, where $S_1^z$ and $S_r^z$ are the $z$-components of a spin
on the cluster boundary and on a rung with distance $r$ to the boundary. In
the upper panel of \fig{fig:magn} this correlation function is shown for two
different paths, where the solid line is $S_{r}$ along one leg of the ladder, and
the dashed line is $S_{r}$ on a zig-zag path through the ladder. Both
correlation functions show clear antiferromagnetic 
correlations along the ladder similar to results obtained by the
finite-temperature Lanczos method.\cite{Aichhorn02} By applying a fictitious
symmetry-breaking magnetic field via the operator $\mathcal{O}(\R)$,
\eq{eq:add_onepartop}, we can estimate whether this ordering is of
long-range type or  not. Similar to \onlref{Dahnken03} we choose for this
field  
\begin{equation}\label{eq:magnfield}
  \Delta_{a,b}=h\delta_{a,b}z_{\sigma}{\rm e}^{i\tilde{\bm Q}{\bm r}_a},
\end{equation}
where $z_{\sigma}$ is $\pm 1$ for spin $\uparrow ,\downarrow$ in the orbitals
$a,b$, $\delta_{a,b}$ is the Kronecker-$\delta$, and $h$ is the field strength.
The wave vector $\tilde{\bm Q}$ is set to $(0,\pi)$ yielding a
staggered field along the ladder. The dependence of $\Omega$ on this
fictitious field is depicted in the lower panel of \fig{fig:magn}. Similar to
the one-dimensional Hubbard model at half filling,\cite{Dahnken03} there is
only one stationary point at $h=0$, which means that the system does
not show long-range antiferromagnetic order, but is rather in a paramagnetic
state with short-range antiferromagnetic correlations.

The above considerations show that the system exhibits short-range
antiferromagnetic spin correlations along the ladder, which can produce the
doubled unit cell. Nevertheless it is also possible that the doubling of the
unit cell is due to short-range charge correlations and not due to spin
correlations. In order to clarify this point we calculated the spectral
function at $V_b=0$ and finite $V_a$, where no charge ordering is
possible. Also in this case the periodicity of the bands with largest
spectral weight is $\pi$ at $k_a=0$ and $2\pi$ at $k_a=\pi$. This shows that
below the phase transition at $V_a=1.3$,
the doubling of the unit cell is mainly due to short-range spin correlations,
and charge correlations play only a minor role in this context.   

When turning to $k_a=\pi$, the spectral function looks totally different. As
on can easily see in \fig{fig:akw_1l_v13} there is hardly any spectral weight
below the chemical potential, which means that there are no occupied states
in the channel $k_a=\pi$. This can be understood, because $k_a=\pi$
corresponds to an antibonding state within a rung, which has energy $2t_a$
relative to the bonding orbital and is therefore not populated in the ground
state.

An obvious difference between the spectra for $k_a=0$ and $k_a=\pi$ is that
in the latter case the excitations with largest spectral weight located between
$\omega\approx 3$ and $\omega\approx 4.5$ disperse with periodicity $2\pi$
instead of $\pi$. Qualitatively this can be understood as follows. When inserting
a particle with $k_a=0$, this electron will occupy a state in the bonding
orbital. Since one of the two states in this orbital is already occupied, the
additional particle must have opposite spin, and thus this particle is
connected to the antiferromagnetic background. A particle with $k_a=\pi$
occupies a state in the antibonding orbital, and since this orbital is not
occupied, both spin directions possess equal possibility. Therefore an
electron with $k_a=\pi$ is not influenced by the antiferromagnetism in the
ground state.

\subsection{Disordered phase including diagonal hopping}\label{subsec:diagonalhop}

So far we have studied single ladders only with hopping parameters $t_a$ and $t_b$
and neglected additional diagonal hopping processes $t_d$, as indicated in
\fig{fig:clusters}. These hopping processes have been important in first-principle
calculations in order to fit the LDA bands correctly.\cite{Yaresko00}
Moreover $t_d$ has been important in C-DMFT calculations in order to describe the
insulating state in the disordered phase properly.\cite{Mazurenko02} In this
section we study the effect of $t_d$ within the V-CPT framework.

\begin{figure}[t]
  \centering
  \includegraphics[width=0.45\textwidth]{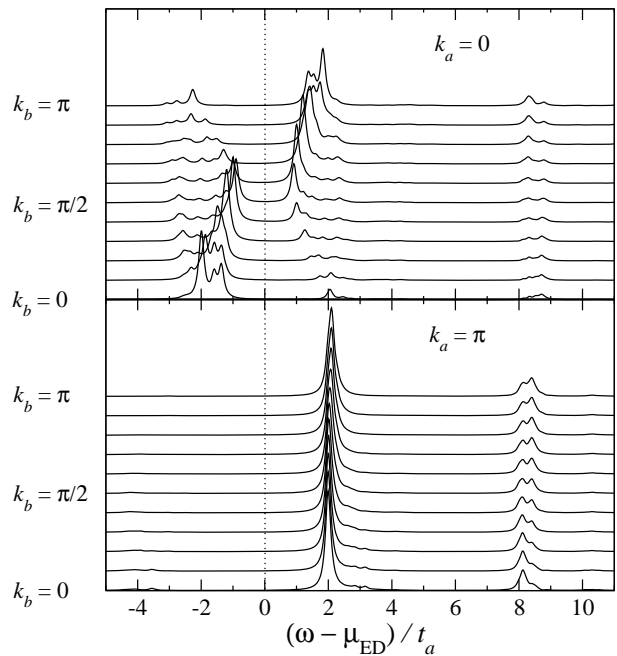}
  \caption{\label{fig:akw_1l_v13_td}%
    Spectral function $A({\bm k},\omega)$ when the diagonal hopping is
    included, $t_b=0.25$, $t_d=0.25$. The Coulomb interaction was
    $V_a=V_b=1.3$. The dotted line marks the chemical potential.}
\end{figure}

In \fig{fig:akw_1l_v13_td} the spectral function is shown for $V_a=V_b=1.3$,
$t_b=0.25$ and $t_d=0.25$, where the hopping parameters are chosen similar
to \onlref{Mazurenko02}. Whereas the spectrum at $k_a=0$ is almost
indistinguishable from \fig{fig:akw_1l_v13}, we see a big difference at 
$k_a=\pi$. There is still hardly any spectral weight below the Fermi energy,
but the band with largest spectral weight above the Fermi level is now located
at approximately $\omega-\mu\approx 2.0$ and can be regarded as dispersionless.

>From a qualitative point of view this can be explained by the dispersion of 
non-interacting fermions on a two-leg ladder in the presence of diagonal
hopping, which is given by
\begin{equation}\label{eq:epsk}
  \begin{split}
  \varepsilon({\bm k}) = &-t_a\cos k_a - 2t_b\cos k_b \\
  &- 2t_d\cos k_a\cos k_b,
  \end{split}
\end{equation}
where the values for $k_a$ are restricted to $0$ and $\pi$, and in these two
cases the dispersion can be written explicitly as
\begin{subequations}
\begin{align}
  \varepsilon(k_a=0,k_b) &=-t_a-2(t_b+t_d)\cos k_b\\
  \varepsilon(k_a=\pi,k_b) &= +t_a-2(t_b-t_d)\cos k_b.
\end{align}
\end{subequations}
This means that for $k_a=0$ the bandwidth is determined by the sum of $t_b$ and
$t_d$, whereas for $k_a=\pi$ it is set by the difference of these two hopping
processes. Since we used $t_b=t_d=0.25$, this fits perfectly to the spectrum
shown in \fig{fig:akw_1l_v13_td}. The sum is equal to the value of $t_b$ used
for the 
calculations without diagonal hopping in \sect{subsec:disordered}, and the
difference is equal to zero, which explains the dispersionless band at
$k_a=\pi$.

The picture that evolves from our calculations is somewhat different to that
obtained in first-principle and C-DMFT calculations. To begin with, the bands
obtained from the LDA all disperse with periodicity $2\pi$ and not $\pi$, as
observed experimentally, along the $b$ direction. Moreover we could not find any
signature of a flattening of the upper $d_{xy}$ bands in the direction
${\bm k}=(0,0)\to(0,\pi)$ when a diagonal hopping is included, which was
reported in Ref.~\onlinecite{Yaresko00}. The main
difference of our calculations to C-DMFT results is that C-DMFT
finds a metal-insulator transition at some finite value of $V$, and this
transition point is shifted downward significantly when $t_d$ is
included.\cite{Mazurenko02}
In contrast we find an insulating state at reasonable values of $U$ already
for $V=0$, even without the inclusion of $t_d$.  The discrepancy to C-DMFT
calculations are very likely  
due to the fact, that the cluster used in the C-DMFT calculations consisted
only of a single rung, and fluctuations along the ladders, which
seem to be important in this system, have been neglected altogether.

\subsection{Ordered phase}\label{subsec:ordered}

We investigate two different driving forces for the occurrence
of a charge-order pattern, (i) the coupling of the electrons to lattice
degrees of freedom, and (ii) nearest-neighbor Coulomb interaction, similar to
\onlref{Aichhorn04}.

Let us start our investigations with possibility (i), the coupling to the
lattice. In order to keep the calculations simple, we consider static
lattice distortions, as discussed in \sect{sec:model}. The inclusion of
dynamical phonon effects would pose a severe problem to the diagonalization
procedures, because for phonons the Hilbert space is a priori of infinite
size, and some truncation scheme has to be applied.\cite{Aichhorn04} 
Well converged results for the spectral function in the presence of dynamical
phonons have so far only be achieved for the polaron and the bipolaron
problem.\cite{Hohenadl03}

\begin{figure*}[t]
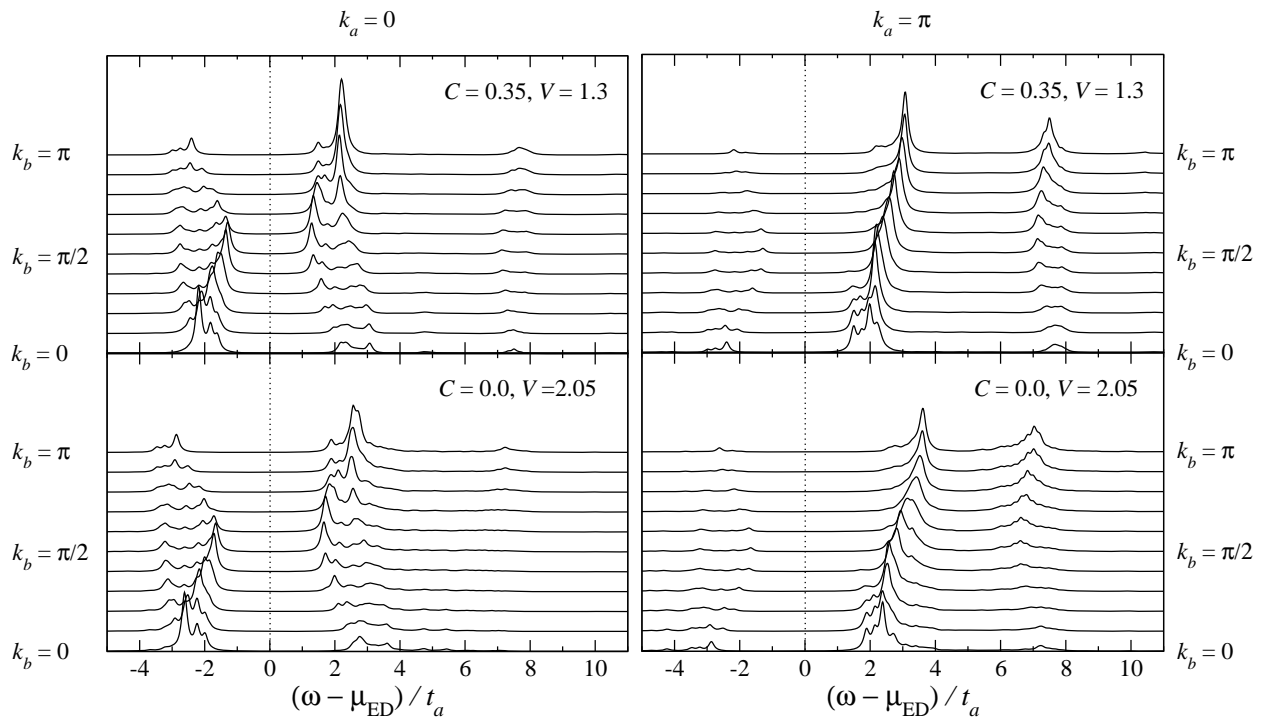

  \centering
  \includegraphics[height=0.4\textheight]{fig8a.eps}
  \includegraphics[height=0.4\textheight]{fig8b.eps}
  \caption{\label{fig:akw_1l_v13_ord1}%
    Spectral function $A({\bm k},\omega)$ in the ordered phase. Top:
    Transition driven by coupling to the lattice. Bottom: Transition driven
    by Coulomb interaction. For the interaction $V=V_a=V_b$ was used, with
    the values as given in the plots. The dotted line marks the chemical potential.}
\end{figure*}

We assumed a zig-zag charge
order pattern, justified by experimental evidence:\cite{Luedecke99}
\begin{equation}\label{eq:zzdist}
    z_i=ze^{i{\bm Q}{\bm r}_i}.
\end{equation}

In order to keep the calculations simple and the number of independent
variables small, we did not consider an additional variational parameter like
a staggered chemical potential. The proper value of the distortion $z$ is
determined as discussed in \sect{sec:model}. 

Motivated by previous work,\cite{Aichhorn04} we set $V_a=V_b=1.3$, since for
this choice we expect the distortions to be close to the experimental value
of $z_{\rm exp}\approx 0.95$ (in units of 0.05\,\AA). Indeed we found
$z=0.911$, which is close to 
$z_{\rm exp}$, and the mean-field parameter was $\delta =0.338$. The spectrum
calculated with these values is shown in the top plots of \fig{fig:akw_1l_v13_ord1}.

The spectral function shows similar features as in the undistorted phase. For
$k_a=0$ the bands disperse with periodicity $\pi$, whereas for $k_a=\pi$ no
evidence for a doubling of the unit cell can be found, and the periodicity is
$2\pi$. Nevertheless, the gap at $\bm k=(0,\pi/2)$ is considerably larger
than for $V=1.3$ without distortions, see \fig{fig:akw_1l_v13}.

An interesting quantity when considering charge-ordering phenomena is the
charge order parameter, which we calculate as
\begin{equation}\label{eq:cdworderpar}
  m_{\rm CDW} = \frac{1}{N_c\langle n\rangle}\sum_j\left( \langle n_j\rangle
  -\langle n\rangle\right)e^{i\bm Q\bm{r}_j},
\end{equation}
where the expectation value $\langle n_j\rangle$ is calculated from the V-CPT
Green's function, and $\langle n\rangle=0.5$.
The factor $\langle n\rangle$ in the denominator assures that the order
parameter is normalized to the interval $[0,1]$. For $V_a=V_b=1.3$ and static
distortions we obtained $m_{\rm CDW}=0.65$, which means that the disproportion of
charges is rather large.

Let us now consider possibility (ii), where the coupling to the lattice is
switched off, $C=0, \kappa=0$, and the charge ordering is driven by the
nearest-neighbor Coulomb interaction. In order to make a connection to the
results obtained with lattice distortions, we calculate the spectral function
at a similar value of the charge order parameter. We found that for
$V_a=V_b=2.05$ the order parameter is $m_{\rm CDW} = 0.66$, close to
the value found above.

The spectral function is shown in the lower plots of
\fig{fig:akw_1l_v13_ord1}. The spectral 
features again look very similar to \fig{fig:akw_1l_v13}. By comparing
the upper and lower plots of \fig{fig:akw_1l_v13_ord1}, one can see that in
both cases the gap at $\bm k=(0,\pi/2)$ is larger than in the disordered
phase, \fig{fig:akw_1l_v13}. To be specific we found a gap size of
approximately $2.6t_a$ in the presence of lattice distortions and $3.4t_a$
without lattice distortions, whereas in the disordered phase the gap was
$1.8t_a$. It is interesting that the momentum-resolved single-particle
spectral features do depend on the driving force of the transition, which
was much less pronounced for, e.g., spin and charge susceptibilities obtained
by integration over the electron states.\cite{Aichhorn04}

The excitation energy near the $\Gamma$ point, relevant for IR experiments,
can be read off from \fig{fig:akw_1l_v13_ord1} to be roughly $4t_a$ with
lattice distortions and $5t_a$ without distortions.
Although these excitation energies are not constant compared to the disordered
phase, calculations including the lattice degrees of freedom 
give a better agreement to experimental IR absorption
data,\cite{Presura00} which show neither a shift of the 1\,eV peak nor
the appearance of new peaks related to electronic transitions.

\section{Results for coupled ladders}\label{sec:results_coupled}

\begin{figure}[t]
  \centering
  \includegraphics[width=0.45\textwidth]{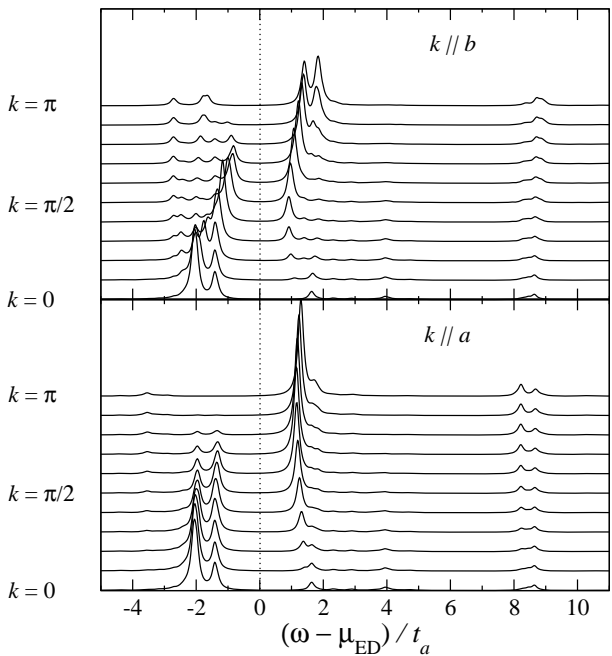}
  \caption{\label{fig:akw_2lad}%
    Spectral function $A({\bm k},\omega)$ in the disordered phase at
    $V_a=V_b=V_{xy}=1.3$ calculated on the $2\times 12$ super-cluster. Top:
    Momentum $k$ along the ladder direction. Bottom: $k$ perpendicular to the
    ladder direction.
\vspace*{1ex} 
} 
\end{figure}

So far we studied only single ladders and neglected the inter-ladder
couplings, since they are frustrated and one might assume that they
are only of minor importance. Nevertheless our approach allows us to include
these inter-ladder couplings by choosing an appropriate cluster geometry, as
indicated on the right hand side of \fig{fig:clusters}. Note that it is necessary
to use a $2\times 12$ super-cluster which allows for a commensurate charge
order pattern across the cluster boundaries. For details of the treatment of
super-clusters we refer the reader to \onlref{Aichhorn04_2}.

The parameter values for the inter-cluster coupling are chosen in the
following way. First-principle calculations have shown that the effective
hopping between different ladders is very small, so we set
$t_{xy}=0.1t_a$, and longer ranging hopping processes are neglected
since the linear dimensions of the cluster are rather small. The values for
the other parameters are the same as used for the calculations in
\sect{subsec:disordered}.

In \fig{fig:akw_2lad} the spectral function for $V_a=V_b=V_{xy}=1.3$ is shown.
For $\bm k$ parallel to the $b$ axis one can easily see
that the spectrum looks very similar to the spectrum of a single ladder
(upper panel of \fig{fig:akw_1l_v13}). The main difference between single and
coupled ladders is that the chemical potential is much larger in the latter
case ($\mu \approx 3.0$), which is due to the frustrated inter-ladder bonds.

When turning to $\bm k$ parallel to the $a$ axis the spectral function
looks very different. The most striking feature is that there is hardly any
dispersion of the bands, and the filled low-energy band can actually be
considered as dispersionless. The spectral weight of this excitation,
however, decreases significantly away from $k=0$ and is transfered
to unoccupied states above the Fermi level at approximately
$\omega-\mu\approx1.5$.

Let us now compare our numerical results to experimental data. Kobayashi {\it
  et.\,al}\cite{Kobayashi98} performed angle-resolved photo-emission
spectroscopy (ARPES) at room temperature, where the system is in the
disordered phase. 
For momentum transfer parallel to the $a$ direction they
found no dispersion of the V $3d$ bands, which fits perfectly well to our
results. For $k$ along the $b$ axis a band dispersion of a 1D
antiferromagnetic quantum system was found with experimental band with
of approximately 0.06--0.12\,eV. This value is rather small compared to the
band width in our calculation of approximately 0.35\,eV, see
\fig{fig:akw_2lad}. We checked that the band width scales with the hopping
along the ladder $t_b$ (not
shown), and therefore this discrepancy between calculations
and experiment can be shortened by choosing a smaller value for $t_b$, which
does not significantly affect the charge ordering of the system.
Nevertheless the strong difference between spectra along $a$ and $b$
direction are well described by our calculations.

\section{Conclusions}\label{sec:conclusion}

In this paper we have applied the recently proposed generalization of the V-CPT for
extended Hubbard models to the case of quarter-filled ladder compounds. We were thus
able to perform the first theoretical study of the spectral function of
$\alpha^\prime$-NaV$_2$O$_5$ within the extended Hubbard model.

For single ladders in the disordered 
phase we found that in
the channel $k_a=0$ the system behaves like a one-dimensional
antiferromagnetic insulator, and the gap is mainly determined by the
nearest-neighbor Coulomb interaction on a rung. Our calculations 
suggest that the system is in an insulating phase for all values of $V$.

This picture still holds when a diagonal hopping $t_d$ is included in the
Hamiltonian, which was suggested to be important by LDA and C-DMFT
studies. We could show that for $k_a=0$ hardly any changes can be seen in
the spectral function, whereas for $k_a=\pi$ the bands become flat. These
findings do not agree with LDA considerations,\cite{Yaresko00}
where a flat upper band was observed for $k_a=0$, thus requiring a finite
value of $t_d$ in a tight-binding fit.

For the transition into the charge-ordered low-temperature phase we considered
two driving mechanisms, the coupling to a static lattice distortion and the nearest
neighbor Coulomb interaction. With lattice coupling we found, similar to
\onlref{Aichhorn04}, that for $V=1.3$ the distortion is close to
the experimentally found size, accompanied by a large 
disproportion of charges. In order to reach the
charge-ordered phase solely by Coulomb interactions, we had to use a
large value of $V$ (with $V=2.05$ for the same value of the order parameter),
which resulted in a large gap in the spectral function, 
considerably larger than in the disordered phase and in the ordered
phase with lattice distortions. Since IR experiments\cite{Presura00} do not
show such a discrepancy, we suggest that for the description
of the ordered phase, lattice distortions cannot be neglected.

Within our approach it was straightforward to study the effects of
inter-ladder coupling on the spectral function. We found that the spectra
along the ladder direction are not significantly affected by these
couplings. Perpendicular to the ladders the calculated bands are almost
dispersionless, in good agreement with experimental data.

\acknowledgments
This work has been supported by the Austrian Science Fund (FWF), project
P15520. M. A. has been supported by a doctoral scholarship program of the Austrian 
Academy of Sciences. We are grateful to E.~Arrigoni and A.~Damascelli for
interesting discussions.


\end{document}